\documentclass[]{spie}  

\usepackage{amsmath,amsfonts,amssymb}
\usepackage{graphicx}
\usepackage[colorlinks=true, allcolors=blue]{hyperref}
\usepackage{multirow}

\title{Challenge of direct imaging of exoplanets within structures: disentangling real signal from point source from background light}

\author[a]{Jialin Li}
\author[a]{Laird M. Close}
\author[a]{Jared R. Males}
\author[a,b]{Sebastiaan Y. Haffert}
\author[c]{Alycia Weinberger}
\author[d]{Katherine Follette}
\author[a]{Kevin Wagner}
\author[a]{Daniel Apai}
\author[e,f]{Ya-Lin Wu}
\author[g]{Joseph D. Long}
\author[h]{Laura Perez}
\author[a]{Logan A. Pearce}
\author[i]{Jay K. Kueny}
\author[i]{Eden A. McEwen}
\author[a]{Kyle Van Gorkom}
\author[a,i,j,k]{Olivier Guyon}
\author[i]{Maggie Y. Kautz}
\author[l]{Alexander D. Hedglen}
\author[a]{Warren B. Foster}
\author[a]{Roz Roberts}
\author[i]{Jennifer Lumbres}
\author[m]{Lauren Schatz}

\affil[a]{Steward Observatory, University of Arizona, 933 N Cherry Ave, Tucson, AZ, USA}
\affil[b]{Leiden Observatory, Leiden University, The Netherlands}
\affil[c]{Earth \& Planets Laboratory, Carnegie Science, 5241 Broad Branch Road NW, Washington, DC, USA}
\affil[d]{Department of Physics and Astronomy, Amherst College, 25 East Drive, Amherst, MA, USA}
\affil[e]{Department of Physics, National Taiwan Normal University, Taipei 116, Taiwan}
\affil[f]{Center of Astronomy and Gravitation, National Taiwan Normal University, Taipei 116, Taiwan}
\affil[g]{Center for Computational Astrophysics, Flatiron Institute, 162 5th Avenue, New York, NY, USA}
\affil[h]{Departamento de Astronomía, Universidad de Chile, Camino El Observatorio 1515, Las Condes, Santiago, Chile}
\affil[i]{Wyant College of Optical Sciences, The University of Arizona, 1630 E University Boulevard, Tucson, AZ, USA}
\affil[j]{Subaru Telescope, National Observatory of Japan, National Institutes of Natural Sciences, 650 N. A'ohoku Place, Hilo, Hawai'i, USA}
\affil[k]{Astrobiology Center, National Institutes of Natural Sciences, 2-21-1 Osawa, Mitaka, Tokyo, Japan}
\affil[l]{Northrop Grumman Corporation, 600 South Hicks Road, Rolling Meadows, IL, USA}
\affil[m]{Starfire Optical Range, Kirtland Air Force Base, Albuquerque, NM, USA}

\begin{document} 
\maketitle

\begin{abstract}
The high contrast and spatial resolution requirements for directly imaging exoplanets requires effective coordination of wavefront control, coronagraphy, observation techniques, and post-processing algorithms. However, even with this suite of tools, identifying and retrieving exoplanet signals embedded in resolved scattered light regions can be extremely challenging due to the increased noise from scattered light off the circumstellar disk and the potential misinterpretation of the true nature of the detected signal. This issue pertains not only to imaging terrestrial planets in habitable zones within zodiacal and exozodiacal emission but also to young planets embedded in circumstellar, transitional, and debris disks. This is particularly true for H$\alpha$ detection of exoplanets in transitional disks. This work delves into recent H$\alpha$ observations of three transitional disks systems with MagAO-X, an extreme adaptive optics system for the 6.5-meter Magellan Clay telescope. We employed angular differential imaging (ADI) and simultaneous spectral differential imaging (SSDI) in combination with KLIP, a PCA algorithm in post-processing, for optimal starlight suppression and quasi-static noise removal. We discuss the challenges in protoplanet identification with MagAO-X in environments rich with scattered and reflected light from disk structures and explore a potential solution for removing noise contributions from real astronomical objects with current observation and post-processing techniques. 
\end{abstract}


\section{Introduction}
\label{sec:intro}  
Protoplanetary disks are the birth sites of stars and planets, thus their structures and chemical compositions provide valuable insights into the underlying mechanisms driving the formation processes of the host stars and their system. High spatial resolution and contrast observation techniques, spanning from millimeter to infrared wavelengths, have revealed a variety of unique disk features, such as spirals, arcs, and gaps. 

It remains unclear whether these substructures can be attributed solely or partially to ongoing planet formation or other processes within the disks. However, interactions between giant protoplanets and their host disks can alter the environment, and the presence of these substructures is often interpreted as a sign of ongoing planet formation \cite{2018ApJ...869L..42H}. Since these larger-scale substructures caused by planetary perturbations are more easily detected than the planets themselves, systems with spiral arms and gaps are targets of interest for high-angular-resolution observations in near-infrared wavelengths. This enables a direct search for protoplanets embedded in disks because they are still radiating remnant energy from the formation process. Although there are reports of such observations, the planetary nature of these companion candidates are ambiguous or challenged (e.g., AB Aur B \cite{2023AJ....166..220Z,2022NatAs...6..751C}; 169142b \cite{2023MNRAS.522L..51H}).

PDS 70 is the only system known to host confirmed protoplanets, both located within its transitional disk cavity. Its inner planet, PSD 70b, was first observed via NIR imaging \cite{2018A&A...617A..44K} and subsequently confirmed through the detection of accretion excess emission in the H$\alpha$ line and in UV continuum \cite{2018ApJ...863L...8W,2021AJ....161..244Z}, and H$\alpha$ differential imaging revealed a second accreting planet in the system, PDS 70 c \cite{2019NatAs...3..749H}. Although other detection of protoplanetary candidates have been reported, they are either located at wider separation or require further study to confirm their planetary nature.  

The lack of confirmed detections can partially attributed to the multitude of challenges involved in directly image exoplanets at small separations. A suite of dedicated hardware, technologies, observing and post-processing techniques are all need to achieve the the high angular resolution at small separations ($\leq$4$\lambda$/D) and high contrast ($\leq$10$^{-3}$) requirements. Due to the complex morphology of transitional disks, observation of these embedded protoplanets faces another challenge: the unambiguous separation of disk and planet signals.

H$\alpha$ differential imaging is one approach to address some of these challenges as it is the strongest recombination line of Hydrogen (H$\alpha$; 656.3 nm) within the optical and infrared wavelengths, that should only come from accreting bodies themselves. When compared to the observed features in IR wavelengths, distinctions between disk structures and protoplanets can be made. With observations in only infrared wavelengths, light from disk structures can be misinterpreted as a protoplanet (i.e. \cite{2017AJ....153..264F}), and the contrast ratio to planet mass relationship in the infrared decreases drastically at lower masses ($<$ 5M$_{Jup}$) making observations of lower mass giant planets more challenging \cite{2014ApJ...781L..30C}. The detection of a point source in H$\alpha$, an emission line tracer for giant planet accretion with a more linear contrast ratio to planet mass relationship, lowers the threshold of detection and allows for a clearer interpretation on the nature of the detection \cite{2014ApJ...781L..30C, 2020AJ....160..221C}.

In this paper, we will discuss recent observations from MaxProtoPlanetS, an H$\alpha$ Protoplanet Survey aiming to discover accreting exoplanets with with MagAO-X, an extreme AO (ExAO) instrument on the 6.5m Magellan Clay telescope. A brief summary of MagAO-X and observational targets are provided in Section \ref{subsec:Obs}. Post-processing procedures for analyzing these highly morphologically complex systems are described in Section \ref{subsec:data} and the results our observations are detailed in Section \ref{sec:results}. We discuss the various reduction methods we used to eliminate noise in each dataset and its similarities to solutions proposed to mitigate exozodiacal dust around planets in the habitable zone in Section \ref{sec:discussion} and the conclusion in \ref{sec:conclusion}.

\section{Observation and Analysis} \label{sec:Observations}
\subsection{Observations with MagAO-X} \label{subsec:Obs}
Observations were made using MagAO-X, a ExAO system designed to perform coronagraphic imaging in visible wavelengths (0.5-1.0 $\mu$m) at high Strehl; it hosts a 97-actuator woofer and 2040-actuator tweeter along with a pyramid wavefront sensor (PWFS) and Lyot-coronagraph system that feeds light into a dual-EMCCD simultaneous differential imaging (SDI) science camera system. The coronagraph contains a third deformable mirror for corrections of non-common path errors. \cite{2022SPIE12185E..09M}. The low noise ($<$0.6 rms e$^{-}$ read noise) EMCCD pyramid WFS OCAM2K detector enables Strehls of $>$50$\%$ while closed loop at 2 kHz. To achieve its science goals of detection and characterization of Solar System-like exoplanets, MagAO-X has recently gone through a Phase II upgrade. Most notably, a new post-AO 1000-actuator MEMS device was added inside the coronagraph to enable Focal Plane Wavefront Sensing (FPWFS) and improved Focus Diversity Phase Retrieval (FDPR \cite{2021JATIS...7c9001V}) performance on sky, increasing the Strehl of $\sim$28$\%$ at H$\alpha$ on sky with faint (V$\sim$12 magnitude) targets \cite{Males2024, Kueny2024}.

With this suite of technologies, the largest and deepest (H$\alpha \sim$10$^{-4}$) survey for protoplanets, MaxProtoPlanetS (PI: Laird Close), has commenced \cite{2020AJ....160..221C,2020SPIE11448E..0UC}. As a part of this project, we observed the following objects in Angular Differential Imaging (SDI) mode with the dual-EMCCD SDI system through an H$\alpha$ continuum filter ($\lambda_o$=0.668 $\mu$m, $\Delta \lambda$\textsubscript{eff}=0.008 $\mu$m) and an H$\alpha$ filter spanning three observation run:

\textbf{(1)HD 34700}, located at 356.5±6.1 parsecs, is a young T Tauri binary with an approximate age of 5 Myr and equal mass components $\sim$2M$_\odot$, is separated by approximately 0.0007$"$. It is known to have three external stellar companions \cite{2004AJ....127.1187T, 2005A&A...434..671S} and exhibits multiple spiral arms along with a massive central cavity with an estimated radius of 175 AU \cite{2009ApJ...690L.110H}.  

The first observation, done on Dec 4 2022, used a wide H$\alpha$ filter($\lambda_o$=0.656$\mu$m, $\Delta\lambda$\textsubscript{eff}=0.009$\mu$m\footnote{Filter specifications can be found in the digital MagAO-X instrument handbook at \url{https://magao-x.org/docs/handbook/index.html}}) along with EM gain of 100 on both science cameras at 4Hz. The V band seeing of the night were affected heavily by the wind, varying from roughly 0.6$"$ to 1.0$"$. As we tracked the object through transit, a total of 51$^\circ$ of sky rotation was obtained. A narrow H$\alpha$ filter($\lambda_o$=0.656 $\mu$m, $\Delta \lambda$ \textsubscript{eff}=0.001 $\mu$m) for the latter epoch of observation on Mar 5 2023. The seeing averaged to be approximately 0.5$"$ throughout the night. An exposure time of 0.25 s, which is equivalent to a 4 Hz readout speed, with EM gains of 50 and 200 were set for science cameras in continuum and H$\alpha$ to avoid saturation. Despite the better conditions, we were only able to obtain about 30$^\circ$ of sky rotation due to its sunset transit time. After data selection, which is detailed in \ref{subsec:data}, we kept 104 min and 132 min of data respectively for the 2022 and 2023 epoch.  

\textbf{(2) HD 142527} is a binary transitional disk system ($\sim$ 5 Myr) located at a distance of 159.3$\pm$0.7 pc. Its primary star is a Herbig Ae/Be star with a mass of $\sim$2 M$_\odot$ \cite{2014ApJ...790...21M}, while it has a lower mass ($\sim$0.35 M$_\odot$) stellar companion situated 15 AU away from the central star \cite{2014ApJ...781L..30C,2016A&A...590A..90L}. Observations in both sub-millimeter and scattered infrared light reveal that the disk of this system exhibits multiple spiral arms and contains a central cavity with a radius of approximately $\sim$140 AU (e.g., \cite{2014ApJ...781...87A, 2021MNRAS.504..782G}). 

This target was observed on the night of Mar 8, 2023 with average seeing of 0.75$"$. The cameras were running at 4 Hz with an EM gain of 40 and 150 respectively through the H$\alpha$ continuum camera and H$\alpha$ narrow camera. We took 160 min of data and acquired 120$^\circ$ of rotation. 

\textbf{(3) MaxProtoPlanetS 1} is a newly imaged ALMA face-on disk with a dust depleted gap at 0.6$"$ with I band magnitude of $\sim$11 \cite{pc1}, placing this object at the faint end of the MaxProtoPlanetS sample. We observed this target on Mar 20 and Mar 25 of 2024 with seeing conditions on both nights being $<$0.5$"$. Again, we utilize the SDI imaging mode through the narrow H$\alpha$ and H$\alpha$ continuum filters. The integration time was set to be 3 seconds per frame for the first observation, and the EM gain was set to 500 for the H$\alpha$ camera and 300 for H$\alpha$ continuum camera. We obtained a total of $\sim$ 70$^\circ$ degrees of rotation for approxmately 70 mins of integration time. Due to its overhead position during transit, 45$^\circ$ of data were before obtained before transit and 25$^\circ$ after transit with 100$^\circ$ rotation spaced between. For our second observation of this object, the EM gain was set to 200 for the H$\alpha$ camera and 600 for H$\alpha$ continuum camera. This latter dataset contained 65 mins of 1s exposure frames after transit, yielding $\sim$ 45$^\circ$ rotation. 

\subsection{Data Selection and Reduction} 
\label{subsec:data}
The selection and reduction procedures are identical for the different science cameras. First, we apply a 10-20$\%$ cut on the dataset by peak counts from the source in each 1024x1024 frame. The center of the star is roughly estimated to be located at the pixel with the maximum counts for the initial selection. A more precise central location is identified in later stages of reduction. For more efficient processing, a box of size 256x256 pixels centered around the central pixel is cropped out from each image. Based on the peak count of the central pixel, frames with values greater than 60,000 counts are rejected as they have surpassed the detector saturation limit.

The remaining frames are aligned via cross-correlation with bi-cubic interpolation through the OpenCV package to account for sub-pixel shifts \cite{opencv_library}. Frames with more than a half pixel offset from the reference PSF are rejected. We note that the sub-pixel offset measurement is done to the precision of the hundredth of a pixel within a 32x32 cutoff around the center; Precision can be increased at the cost of longer run time. An additional $\sim$20-30$\%$ of frames are discarded for their large offset for most datasets. With worse data quality (i.e. low Stehl ratio, increased seeing, etc.), the percentage can increase to roughly 50$\%$. 

Since the expected H$\alpha$ flux from the protoplanets is low, it is key to preserve every photo-electron from the planet to ensure the survival of planet light after PSF subtraction. The passing frames are block averaged by time or parallactic angle (PA) for better performance of the pipeline. The former method combines a certain numbers of frames into one via summation, creating a combined frame with a longer exposure time and larger number of H$\alpha$ counts. Such number varies depending on the size of the dataset and we have found that constraining this parameter somewhere between 50 to 200 frames leads to higher efficiency and ensure the diversity in PA for increased ADI performance. The PA value of each combined frame is the average of the PA values associated with each individual frame. The latter is an alternative method for observations without data through transit, as it creates a more evenly spaced data cube to enhance ADI performance. 

The final step before PSF subtraction is applying a high-pass filter to the combined frames, which serves as a substitute for dark and flat subtraction. Additionally, it also removes some of the extensive scattered light from the central star and dust present in the system which occupies the lower spatial frequency space. A median image of the combined frames is created to determine the true center of the aligned images, the FWHM and peak of the PSF. 

We used the python implementation package PyKLIP \cite{2015ascl.soft06001W} to perform both KLIP and ADI to increases the post-processed contrasts. There are three key parameters in PyKLIP that effects the stellar PSF reconstruction and subtraction: annuli, subsection, and numbasis. The PSF is modeled in annular segments (${\tt annuli}$), each subdivided into equal subsections (${\tt subsections}$) and with a number KL modes or principal components (${\tt numbasis}$). PyKLIP produces a de-rotated data cube containing the images after the inputted number of principal components removed. In our case, our post KLIP-ADI data cube contains images with 1, 5, 10, 20 and 50 KL modes subtracted. The movement parameter, an exclusion criteria for picking reference PSFs, were limited to values ranging from 0-5 pixels. In other words, the reconstruction of the target PSF does not utilize images where the rotation of the companion between target and reference is less than the given movement value. If the movement values are small, the final reduced images can be more susceptible to self-subtraction. 

The final ASDI images are created by multiplying the KLIP-ADI reduced continuum image by a scale factor before subtracting it from the H$\alpha$ image to account for the flux difference from primary star at the two wavelengths. This should eliminate the residual starlight and scattered light from disk structures. In the case where the disk structures are not located close to the point source of interest, the scale factor is a ratio of flux of the central star between the two filters. To account for the change in diffraction pattern with wavelength, the continuum image is scaled spatially by the ratio of the two wavelengths. However, in the case where a bright disk structure is present, this step of resizing the continuum image is omitted and the scaling factor for flux is determined through minimizing the average flux difference of apertures placed on disk structure around the region of interest between the two wavelengths for each mode parameter. Further discussion on optimization of the free parameters of the different reduction routines used for datasets and their effects on the interpretation of the results can be found in Section \ref{sec:discussion}. 


\subsection{Astrometry and Photometry of Companions Candidates} \label{subsec:astro+contrast}
As PSF subtraction algorithms can distort planet signal, we obtained companion astrometry and photometry through the Bayesian KLIP Astrometry (BKA) technique with the forward modeling feature in PyKLIP \cite{2016AJ....152...97W, 2016ApJ...824..117P, 2015ascl.soft06001W, 2013PASP..125..306F} for accurate measurements and uncertainties on the companion parameters. The initial position of the planet is obtained through a Gaussian fit in ASDI images, and a grid of forward modeled negative planets are injected within a FWHM of such position into the datacube of combined frames. The injected planets are all of the same brightness and thus minimizing the total flux within a circular aperture with radius being FWHM of the PSF centered around the initial position should be the optimal location. To account for the uncertainty of the initial position, the optimal location is taken to be the median value of positions with minimized total flux with aperture radii ranging from 0.75 FWHM to 1.5 FWHM. The contrast of the companion can be determined in a similar method through injecting negative fake planets of various contrast at the optimized position, but we minimize the root-mean-square (RMS) rather than total flux. 

\section{Results} \label{sec:results}

\begin{figure}[ht!]
\centering
\includegraphics[width=0.82\textwidth]{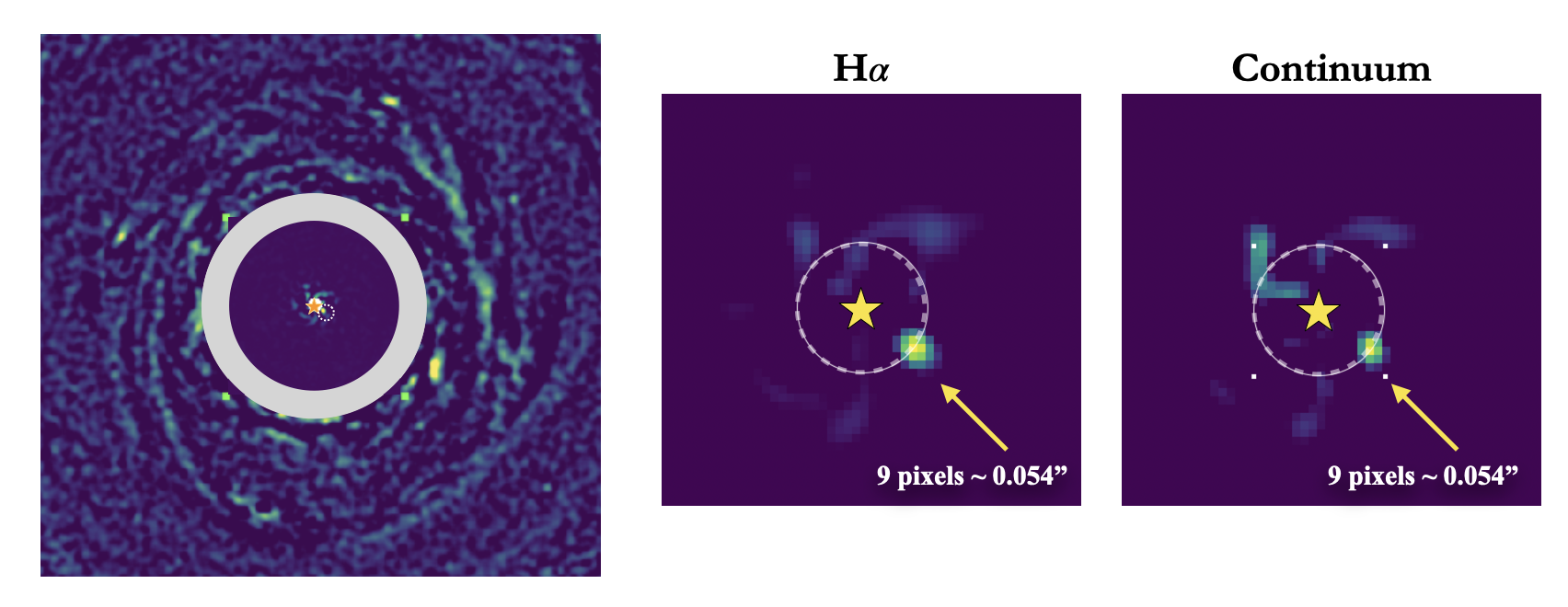}
\caption{The composite reduced KLIP (movement=1, numbasis=5, High Pass (HP) filter width=4.5 pixels) in H$\alpha$ continuum showcasing the multi-spiral Disk with the full frame FOV of $\sim$2.5$"$ by 2.5$"$. Note north is up and east is left. The grey annulus indicates the change in stretch in this image. The stellar companion, HD 142527B, can be seen more clearly in the H$\alpha$ and continuum images respectively in the middle and right panels. The circle aperture marks a 3 pixel radius around the location of the stellar binary. The excess H$\alpha$ flux from HD 142527B is clear and the continuum flux is likely emission from the photosphere of this low mass M star.}
\label{fig:HD142527}
\end{figure}

\subsection{Objects with Known Companions or Companion Candidates in the Literature}
\textbf{HD 142527:} Shown in Figure \ref{fig:HD142527}, we successfully resolved the stellar binary of HD 142527 with a SNR=5 at $\sim $0.054$"$ ($\sim$5 AU) in both science filters, as well as the disk spanning 1$"$ in radius. There are no additional detections of new companion candidates in the gaps of this system.  

\subsection{Objects with New Companion Candidates}
\begin{figure}[ht!]
\centering
\includegraphics[width=0.82\textwidth]{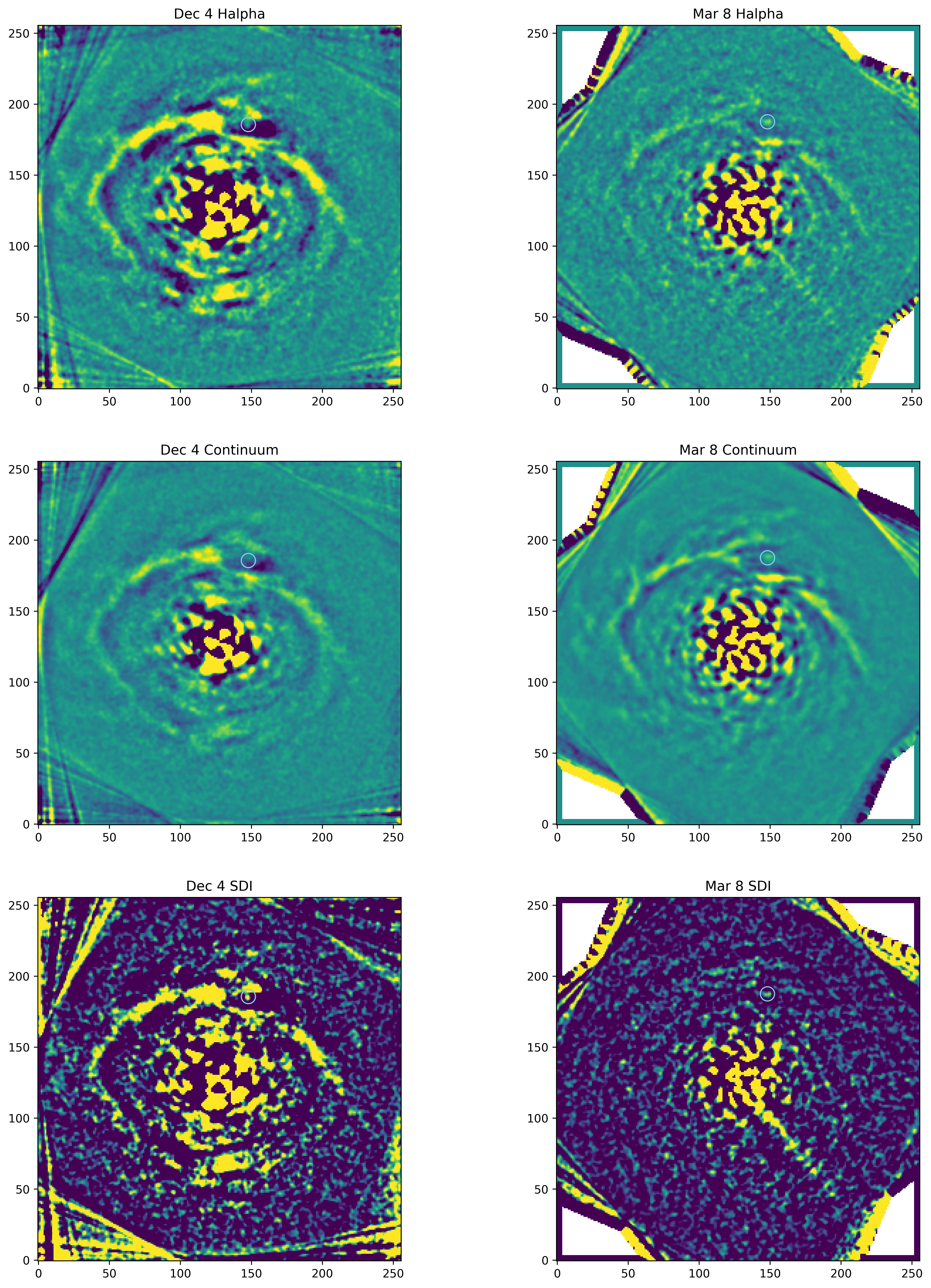}
\caption{The reduced KLIP (movement=2, numbasis=1, HP filter width=4.5) with additional smoothing of the H$\alpha$, continuum, and ASDI images from Dec 4 and Mar 8 are respectively on the left and right columns. Note north is up and east is left.  An additional 1-pixel wide Gaussian kernel is convolved with the reduced KLIP image to produce these images. The circle aperture marks a 5 pixel radius around the location of the companion candidate.}
\label{fig:SDI}
\end{figure}

\textbf{HD 34700:} We report the tentative ($\sim$ 4$\sigma$) detection of a point source with excess H$\alpha$ flux in both the Dec 4 and Mar 8 observations. The reduced H$\alpha$, continuum images and final SDI cubes from both epochs are shown in Figure \ref{fig:SDI}. The brightness of the object and its massive disk introduces light contamination in the position of the observed point source and the region around it (represented by the circular aperture in Figure \ref{fig:SDI} and beyond). In the first epoch, the point source appears to be an extended structure of the disk in both the H$\alpha$ and final ASDI image. Although a similarly prominent source was lacking in the continuum images at the position of the structure excess H$\alpha$ flux, we do note that there is a significant amount of background light either from the scattered disk light or remaining flux from the stellar PSF. If treating this source as a potential companion, it is an 5.3 $\sigma$ point source detection at separation of 61.0$\pm$1.27 pixels and PA of 341.0$\pm$0.72$^{\circ}$, with a contrast of (5.5$\pm$2.7)$\cdot$10$^{-5}$. 

In the latter epoch, with better seeing conditions and a narrow H$\alpha$, there is an significant reduction of residual stellar flux and scattered light overall. Finer substructure of the disk can be identified in the continuum, including western parts of the inner disk. Due to the lack of photons through the narrow 1 nm H$\alpha$ filter, the path of the central photo-electrons being transferred across detector during readout becomes prominent (PA$\sim$ 142$^{\circ}$), leaving a similar linear ``read-out stripe" structure in the ASDI images. The H$\alpha$ excess source appears to be detached from the disk and has a closer resemblance to a point source in both H$\alpha$ and continuum KLIP reduced images. As the value of the numbasis parameter increases , or the number of KL basis used for stellar PSF reconstruction and subtraction, the bright point source in continuum images changes into a dipole like structure, with two bright circular blobs with a dark lane separating the two in the middle. When treating this source as a potential companion, the point source is located at a separation of 62.9$\pm$0.87 pixels and PA of 341.3$\pm$0.57$^{\circ}$, with a contrast of 5.5$\cdot$10$^{-5}\pm$2.7$\cdot$10$^{-5}$ and an average SNR of 4.

\begin{figure}[ht!]
\centering
\includegraphics[width=0.7\textwidth]{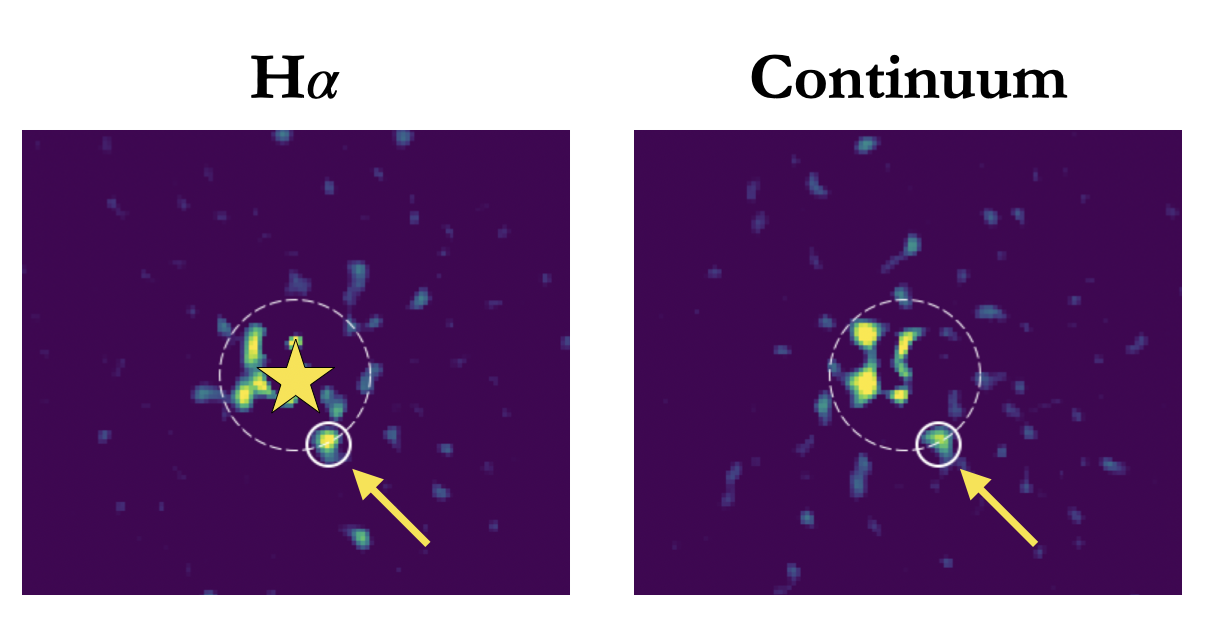}
\caption{The reduced KLIP (movement=1, numbasis=5, HP filter width=7) with additional smoothing of the H$\alpha$ and continuum. This reduction uses combined frames binned by PA. An additional 1 pixel wide Gaussian kernel is convolved with the reduced KLIP image to produce these images. The circle aperture marks a 3 pixel radius around the location of the companion candidate.}
\label{fig:2mj-sdi}
\end{figure}

\textbf{MaxProtoPlanetS 1:} We report a detection of a companion candidate in the second epoch (Mar 25) observation of this object in both H$\alpha$ and continuum images at 5 $\sigma$ and 2 $\sigma$ respectively. The H$\alpha$, continuum, and ASDI images are shown in Figure \ref{fig:2mj-sdi}. Despite the positive detection in both filters, there appears to be a $\sim$1 pixel offset in the location of the source and a difference in morphology. The H$\alpha$ source resembles a point source in different reductions more consistently, which is in agreement with a true accreting object \cite{2020AJ....160..221C}.

It is also worth noting that we fail to obtain a 5$\sigma$ source when choosing to combine frames by time rather than PA, like attributable to the lack of significant rotation for many of the frames taken well after transit, as mentioned in \ref{subsec:data}. The 5$\sigma$ detection in ASDI image indicates the presence of additional H$\alpha$ flux from the source. 

However, we fail to detect any signal in the initial observation on Mar 20 in both wavelengths with either method of performing ADI as shown in the top row of Figure \ref{fig:2mj-epoch1}. Considering the variations in the PSF before and after transit, we discarded the data post-transit ($\sim$15$\%$) for a more stable PSF reference to ensure the efficiency of PSF removal with PyKLIP. The reduced pre-transit dataset is shown in the bottom row of Figure \ref{fig:2mj-epoch1}. 

\section{Discussion} \label{sec:discussion}

\textbf{HD 142527:} The location of the stellar companion is roughly consistent with the latest orbit with semimajor axis $\sim$10 AU \cite{2024A&A...683A...6N}. Our additional data point will further constrain the orbit of HD 142527 B, but the separation remains small between the central binary and thus ruling out the highly eccentric orbit of the stellar companion as the cause to many of its disk features, notably the massive cavity spanning $\sim$100 AU. With the non-detection of additional companions in this system, the origin to its large central cavity remains unknown, as the stellar companion with such an orbit can only be responsible for gaps $\sim$ 30 AU \cite{2024A&A...683A...6N}. 

\textbf{HD 34700:} The positive H$\alpha$ source identified in both epoch of MagAO-X observations is approximately 0.378$"$ away from the central binary. A SNR=5 source with H$\alpha$ excess emission was detected in the first epoch. However, due to sub-optimal seeing conditions and the brightness of the disk in H$\alpha$, it was impossible to separate the contributions from the disk and the planet signal as the signal does not appear to be a point source, but an extension of the disk. In subsequent observations in the 2023A term, we employed narrow H$\alpha$ filter ($\Delta\lambda$= 1 nm) instead of the regular H$\alpha$ filter ($\Delta\lambda$= 9 nm). This allowed for the isolation and recovery of the excess H$\alpha$ signal from the protoplanet candidate, albeit with an SNR= 3.5. The improved conditions during this epoch also revealed a positive signal in the continuum emission at the same location. Although the continuum signal did not resemble a point source as observed in H$\alpha$, the presence of continuum flux suggests that the source may be associated with disk features rather than an accreting protoplanet. 

\begin{figure}[ht!]
\centering
\includegraphics[width=\textwidth]{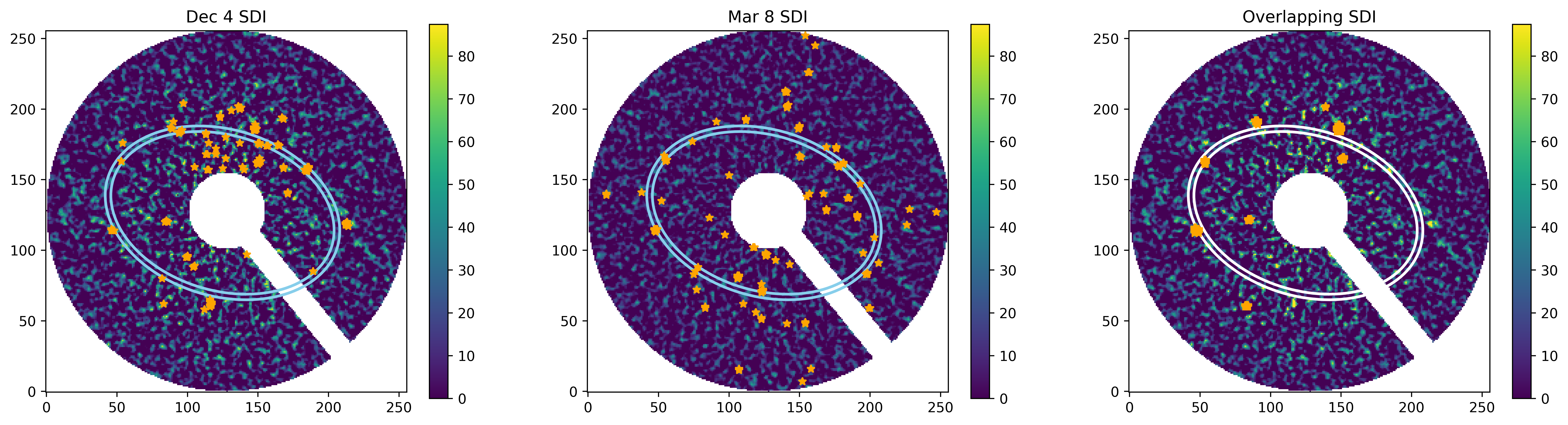}
\caption{The ASDI images from 2 epochs of HD 34700 observations are shown respectively on the left and middle panels with north up and east on the left. The panel on the right shows the overlap of positive values between 2 epochs. The orange dots represents the brightest pixels within the images and the elliptical annulus represents the rough location and thickness of the disk. Brighter pixels from the star, artifacts from the edge of the image, and the ``readout-stripe" are masked.}
\label{fig:overlap_SDI}
\end{figure}

When combining both SDI images and overlapping apertures of a five pixel radius centered on the 100 brightest pixels from each epoch, only a handful apertures remain as shown in Figure \ref{fig:overlap_SDI}. Majority of remaining apertures are located on the brighter parts of the disk, and only two does not have direct association with the disk, which is illustrated roughly by the gray elliptical annulus in Figure \ref{fig:overlap_SDI}. One of such two location appears to be consistent with the location of the potential companion. 

Observation in JHK bands with SCExAO/CHARIS detected a positive signal with close proximity to our observation \cite{2020ApJ...900..135U}. However, due to the complex disk structure, they concluded that the signal is likely a distorted part of the disk's spiral arm or an artifact introduced during post-processing. Mass limits on the potential substellar mass objects were placed at $\sim$12 M$_{Jup}$ inside the disk and $\sim$5 M$_{Jup}$ outside of the disk. The true nature of this source remains a mystery, but its excess of H$\alpha$ flux is challenging to explain as anything other than a protoplanet. 

A recent multi-wavelength study of this disk revealed an inner ring extending from 65 to 120 AU inside the multi-spiral outer disk in the polarized H$\alpha$ image\cite{2024A&A...681A..19C}. We detected parts of this inner disk feature in both epochs, but as it is an extended structure, it is broken up into sections by KLIP, and is better seen in the latter epoch of data. Although there were no detection of point sources, they used the observed geometric offsets between the inner and outer ring to constrain the mass and location of the potential companion to be a $\sim$4$_{Jup}$ mass planet inside or outside the HD 34700 A inner disk \cite{2024A&A...681A..19C}. Due to the difference in filter width in the new epoch, we failed to remove the speckle noise and identify any protoplanetary candidates within the inner disk. 

\begin{figure}[ht!]
\centering
\includegraphics[width=\textwidth]{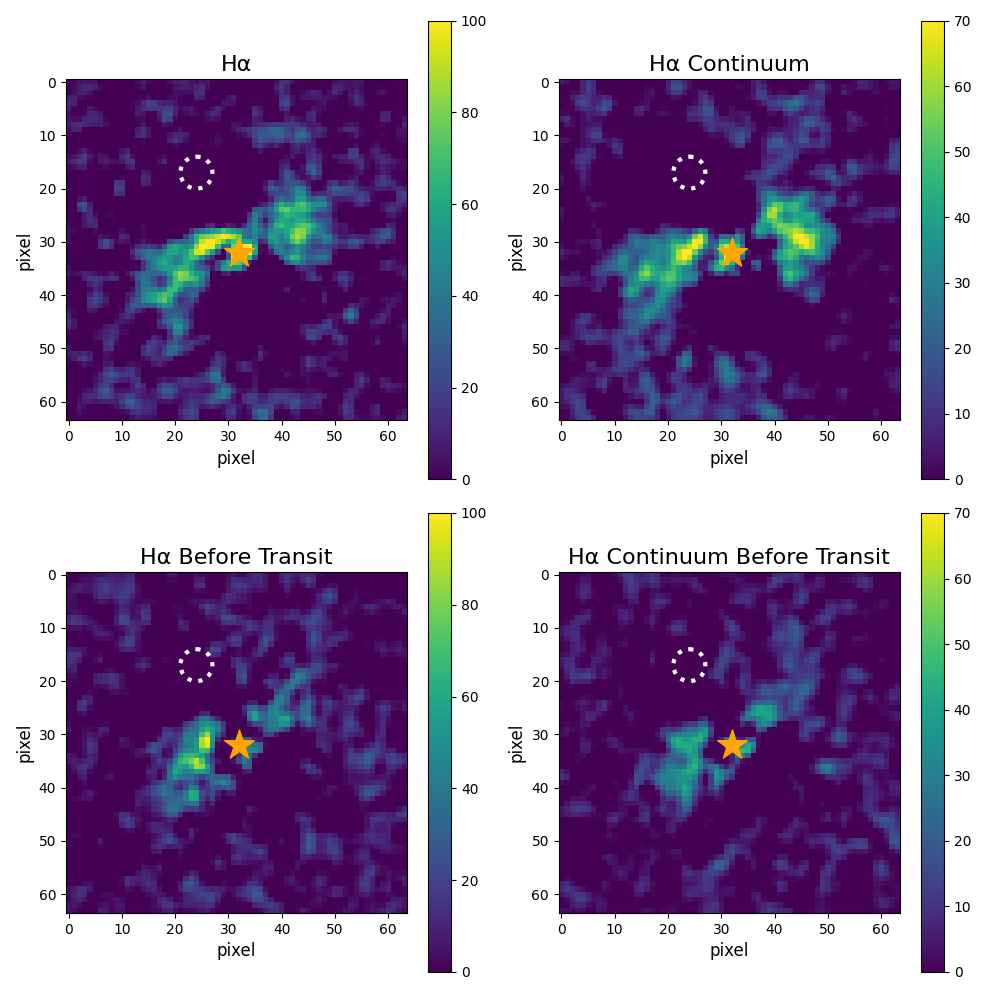}
\caption{The reduced KLIP (movement=1, numbasis=5, HP filter width=10) with additional smoothing of the H$\alpha$ and continuum images respectively on the left and right columns of MaxProtoPlanetS 1 in the first epoch. The top row are reductions with all selected data while the bottom row are reductions of data before transit. Note that the combined frames are stacked by their PA angle rather than time. An additional 1 pixel wide Gaussian kernel is convolved with the reduced KLIP image to produce these images. The circle aperture marks a 3 pixel radius around the location of the companion candidate from the second epoch of observation, however, only wind driven halos can be seen.}
\label{fig:2mj-epoch1}
\end{figure}

\textbf{MaxProtoPlanetS 1:} Our null detection in the first epoch of observation of this object is likely attributable to the fast variation in atmospheric turbulence conditions that was unable to be corrected by the AO loop. We see strong wind driven halos and its artifacts in the post-processes images. Through decreasing the variation in the PSF by eliminating data obtained after transit, the wind driven halo becomes less prominent as shown in Figure \ref{fig:2mj-epoch1}. However, the smaller variation still prevents the ADI from accurately modeling the starlight residuals, resulting in strong asymmetric wind driven halo residuals the ADI images. Additionally, since the direction of the wind driven halo follows PA, which is the same PA as the companions, its signature aligns and accumulates when the temporal data cube is simply rotated and median combined \cite{2020A&A...638A..98C}. This behavior we observe is likely caused by the smearing of the servo-lag speckles across a planet in the direction of a wind. 


\subsection{``Read Out Stripe$"$} \label{subsec:readout}
One of the major sources of noise in the post-processed images is a bright stripe created from the transfer of charge from the bright central star to the readout region on a frame transfer CCD. An example in our data can be seen in as the linear feature the Mar 8 data of HD 34700A in the lower right panel of Figure \ref{fig:SDI}, where this effect is most prominent when the camera is set with high EM gains and high readout speeds on targets with low photon counts. As HD 34700A and its disk are brighter in continuum and the continuum EM gain is relatively lower than its H$\alpha$ counterpart, the stripe is not pronounced in the H$\alpha$ continuum ADI image. 

The stripe remains in the same position throughout observations, just like stellar speckles, thus it can be partially removed by ADI PSF subtraction and its effect is negligible. However, when the observation lacks rotation, it remains and gets broken up by PyKLIP, resulting in bright blobs or point sources in that PA vicinity. The location of the stripe is flipped between the cameras due to the extra reflection towards the H$\alpha$ science camera, allowing for partial differentiation between artifact and true signal. For protoplanet identification, which requires differentiation in both cameras to distinguish its nature, we neglect the regions overlapping with the read out stripe.

\subsection{Disk Removal in ASDI} \label{subsec:contamination}
In attempts to isolate the point source signal in HD 34700, we used a slightly different approach creating the ASDI images. As mentioned in Section \ref{subsec:data}, we did not compress the continuum image by the ratio of the wavelengths for this object, as this standard practice of SDI only enhances our major noise source, disk structures; Unlike diffraction speckles, real objects like disks, do not scale with wavelength. As the source is located within the disk, the disk must be removed to isolate the point source. Due to the large size of disk and its distance from its young central stars, the flux from the central star no longer accurately represents the flux of the disk. Thus, we attempted to remove the outer disk via two different approaches. 

\begin{figure}[ht!]
\centering
\includegraphics[width=0.75\textwidth]{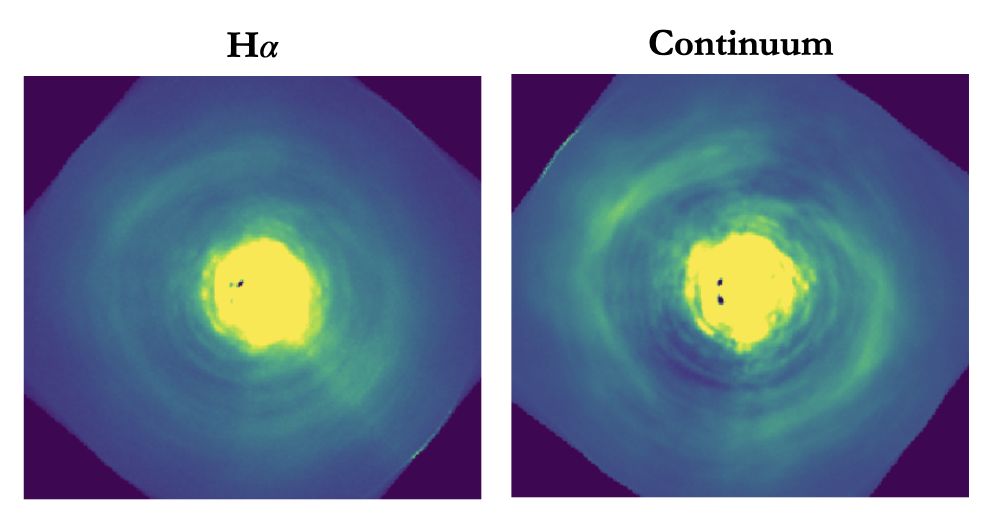}
\caption{The reduced image of HD 34700A of the 2023A epoch with KLIP (movement=1, numbasis=5, HP filter width=4.5), ADI, and RDI applied in both images. The main outer disk structures are extracted while minimizing KLIP self-subraction artifacts. Note north is up and east is left.}
\label{fig:rdi}
\end{figure}

First, we first performed reference differential imaging (RDI) in combination with KLIP to extract the outer disk in observations in both wavelengths with a bright target observed earlier in the night as the reference star. The extracted disk can then be subsequently subtracted from the ADI image. The outer disk can be extracted using this method as shown in Figure \ref{fig:rdi}. While the rough elliptical shape of the outer disk is extracted, its multi-spiral and discontinuity feature cannot be recovered. Due to the lack of the disk structures in the model, the disk removal only added more noise to the region of interest near the discontinuity.

We tried removed the disk by performing photometry on the disk as an alternative; We measured the total flux from the outer disk via a single elliptical annulus and a group of circular apertures in both wavelengths for creating the final ASDI image for both epochs. The elliptical annulus is fitted to the outer disk with the spirals ignored via the Least Squares fitting of ellipses tool in Python \cite{ben_hammel_2020_3723294}. The slight inclination of the outer disk introduced a variation of brightness in disk, causing the region of interest to have higher flux than the rest of the disk. The scaling factor determined through a single elliptical annulus represents a more global flux different of the disk, but is not effective in removing the disk in region of interest. To find a scaling factor better representative of the residual disk flux in the northwest, we placed a group of circular aperture with diameter of size equivalent to the FWHM of the PSF in such region. We experimented with different numbers of apertures placed and their locations, and found that placing eight on the bright discontinuity region of the disk can best removes the disk around the companion candidate. This is likely due to the flux from the structure around the discontinuity better captures the true flux within the region of interest.

However, this elongates the candidate and creating a more extended morphology rather than a point source. Furthermore, this approach to disk removal introduces new variables when performing photometry of the disk, such as the location, shape, size, and number of the apertures used. Fine-tuning these parameters can lead to different interpretation on the properties and the nature of the detected source in the ASDI image.  

The task of retrieving the planet from disk structures shares similar challenges as imaging planets within habitable zones. The exozodiacal dust present in the region can obscure observations of Earth-like exoplanets and it will need to be subtracted to achieve the desired contrast.  The various techniques used in the proposed solutions are also utilized in our reduction pipeline: PSF subtraction, ADI, and disk subtraction via high pass filtering \cite{2022AJ....164..235K,2023AJ....166..197C}. In simulations, a simple high-pass filter removes structured exozodi to the Poisson noise limit for systems with inclinations  $<$60$^\circ$ and up to 100 zodis \cite{2023AJ....166..197C}. However, to reach such a noise floor in real observational data can be challenging, especially when there are bright dust or speckle structures at the same spatial frequencies as the planet. Our observations of the HD 34700A system serve as a realistic example in non-ideal scenarios, including the presence of the bright and complex disk structures and observational conditions being sub-optimal.

\section{Conclusion} \label{sec:conclusion}
In this paper, we presented MagAO-X observations of three different transitional disk systems on the 6.5m Magellan Clay telescope using H$\alpha$ differential imaging for extraction of accretion emissions from potential protoplanets. We applied standard PSF subtraction and post-processing techniques to separate planetary signals from the speckle noise and the surrounding disk features. We successfully identified accreting gap companion signal in all three datasets:  a repeated detection of the accreting stellar companion in the HD 142527 system along with two potential substellar companion candidates, HD 34700Ab and MaxProtoPlanetS1. 

Despite the fact that both protoplanet candidates had 5$\sigma$ detections in a single epoch, the true nature of these two objects remains a question due to observation variability across epochs and the disk background noise they are embedded in. With our current observational and post-processing techniques, we find that the wind driven halo can limit our sensitivity in region closer to the star and complex disk features can become a problem as we move further away. We do not have a reliable approach to disentangle embedded protoplanet light from its disk that doesn't appear as a clear H$\alpha$ excess point source. However, our dataset of HD 34700A serves as an example of an embedded source that can be recovered, but the nature of the source is enigmatic and our achieved contrast is reduced due to disk light. This can be a similar problem faced when trying to imaging habitable zones planets within exozodiacal dust.

\acknowledgments 
J. Li, M. Y. Kautz, and E. A. McEwen are supported by NSF Graduate Research Fellowships. L. M. Close and J Li. were partially supported by NASA eXoplanet Research Program (XRP) grant 80NSSC18K0441 and is now supported by grant 80NSSC21K0397, which funds the MaxProtoPlanetS survey (PI: L. M. Close). S. Y. Haffert received support from NASA through the NASA Hubble Fellowship grant \#HST-HF2-51436.001-A, awarded by the Space Telescope Science Institute (operated by AURA under NASA contract NAS5-26555). We are very grateful for the support from NSF MRI Award \#1625441 for the development of MagAO-X. The MagAO-X Phase II upgrade program (PI: J. R. Males) is made possible by the generous support of the Heising-Simons Foundation.

\bibliographystyle{spiebib} 
\bibliography{main} 

\end{document}